\documentclass[a4paper,aps,pra,showpacs,preprintnumbers]{revtex4}
\usepackage{graphics,graphicx,epsfig}
\usepackage[centertags]{amsmath}
\usepackage{amsfonts}
\usepackage{amssymb}
\usepackage[english]{babel}
\usepackage{graphicx}
\usepackage{amsthm,color}

\begin{document}

%%%%%%%%%%%%%%%%%%%%%%%%%%%%%% User specified LaTeX commands.

%\renewcommand{\vec}[1]{\mathbf{#1}}

\def\BE{\begin{equation}}
\def\EE{\end{equation}}
\def\BY{\begin{eqnarray}}
\def\BEA{\begin{eqnarray}}\def\EY{\end{eqnarray}}\def\EEA{\end{eqnarray}}
\def\L{\label}
\def\nn{\nonumber}
\def\({\left (}
\def\){\right)}
\def\<{\langle}
\def\>{\rangle}
\def\[{\left [}
\def\]{\right]}
\def\o{\overline}
\def\BA{\begin{array}}
\def\EA{\end{array}}
\def\ds{\displaystyle}
\def\c{^\prime}
\def\cc{^{\prime\prime}}

\title{Entanglement measurement of the quadrature components without the homodyne detection
in the spatially multi-mode far-field }
\author{ T. Golubeva, Yu. Golubev,  K. Samburskaya,
}
\address{V.~A.~Fock Physics Institute, St.~Petersburg State University,\\ 198504 Stary Petershof, St.~Petersburg,
Russia}

\author{C. Fabre, N. Treps}
\address{Laboratoire Kastler Brossel, Universit\'{e} Pierre et Marie Curie-Paris6, Place Jussieu, CC74, 75252 Paris
Cedex 05, France}

\author{ M.~Kolobov}
\address{Laboratoire PhLAM, Universit\'e de Lille 1,
F-59655 Villeneuve d'Ascq Cedex, France}
\date{\today}

\begin{abstract}
We consider the measuring procedure that in principle allows  to
avoid the homodyne detection for the simultaneous selection of
both quadrature components in the far-field. The scheme is based
on the use of the coherent sources of the non-classical light. The
possibilities of the procedure are illustrated on the basis of the
use of pixellised sources, where the phase-locked sub-Poissonian
lasers or the degenerate optical parametric oscillator generating
above threshold are chosen as the pixels. The theory of the
pixellised source of the spatio-temporal squeezed light is
elaborated as a part of this investigation.
\end{abstract}

\maketitle

\section{Introduction}
The subject of quantum optics always implies the use of some kind
of nonclassical light. Therefore under the conditions of  the real
experiment one of the basic goals is to find a comprehensible
source that gives nonclassical light with acceptable
spatio-temporal properties. For now  there are several kinds of
sources such as optical parametric amplifiers and oscillators
generating below threshold (see, e.g., \cite {1}) that are
non-alternatively used in the practice. These devices are
preferable mostly because of  the nature of parametric interaction
itself, which automatically ensures a generation of the squeezed
light states \cite {Lugiato1,Lugiato2,Lugiato3,Lantz}.

Together with the obvious advantages of the parametrical sources their applications meet the well-known difficulties.
One of them, for example, is a necessity of the balanced homodyne detection use. In the experimental practice a
preparation of the local oscillator with the acceptable spatio-temporal configuration can be not a simple operation
especially for quantum imaging. So it is reasonable to consider the other alternatives. In this article, the light
sources are considered to be with quantum fluctuations occurring on high coherent level. We believe that the similar
sources could provide us with new possibilities in comparison with traditional ones. This can be realized by  devices
such as lasers \cite{2,3,4} or by any parametric device generating above threshold \cite{5,6}. These sources we will
call \emph{coherent sources}. As we shall demonstrate further they allow us to avoid an obligative use of the
homodyne detection procedure, since the coherent component of the radiation itself constitutes for the local
oscillator amplitude.

The development of quantum optics during the last decade has clearly demonstrated that for reasons of any
applications the greater number of degrees of freedom in the non-classical radiation is a positive and very important
factor allowing to essentially improve an efficiency of optical measuring procedures. From this standpoint the use of
the single-mode coherent sources does not seem to be an attractive perspective. However as it was demonstrated in
Ref.~\cite{2} such conclusion is too hasty because even in this case the radiation outside the high-Q cavity in free
space turns out to be broadband.

Having the good temporal statistics the single-mode lasers or the optical parametric oscillation are often ill
acceptable for the analysis of spatial structures, for example, for the aims of quantum imaging. To adapt them to the
spatially multi-mode schemes it is suggested to use the so-called pixellised sources that were discussed in  Refs.
\cite{3,5}. These sources are formed by putting periodically the pixels (lasers or optical parametric oscillators) on
a plane surface. That provides us with a possibility to effectively use both the spatial and temporal degrees of
freedom in distinguish from the optical parametric oscillator generating below threshold that was in particular  used
in the dense-coding protocol \cite{7}, where the temporal degrees freedom could not be enable.

In this article we use the following  abbreviations: coherent source (CS); homodyne detection (HD); pixellised source
(PS); degenerate optical parametric oscillator (DOPO); sub-Poissonian laser (SPL).

 This article is organized as follows. In Secs.~\ref{II} and \ref{III}, two  ways (with one and
 with two sources and correspondingly with one and two detectors )
  of a direct detection  of the quadrature components are considered. In Secs.~\ref{IV},\ref{V}, and \ref{VI} the measurement
  procedure is applied for the specific sources, namely for the isolated pointlike phase-locked SPL and DOPO,
  for the PSs on the  basis of the SPL and DOPO.

\section{Direct detection of the quadrature components in the far-field (scheme with single coherent source) \L{II}}
In this article our main aims are coupled with the problems of
quantum optics and quantum information that is the squeezed and
entangled light states are specially important for us. However we
would like to emphasize that the results obtained in
Secs.~\ref{II} and \ref{III} and devoted to the selection of the
quadrature components without the use of the homodyne detection
technique can have more wide application and are not directly
coupled with any quantum aspects in the field.

\begin{figure}
% \centering
\includegraphics[height=3cm]{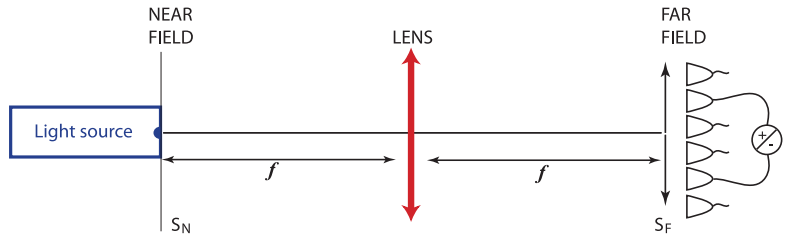} \caption{Illustration of the near- and far-field, $f$ is a focal distance,
 $z$ is a direction of light propagation.}
 \label{fig1}        % Give a unique label
\end{figure}

We shall consider two ways  to select the quadrature components presented correspondingly in Fig.\ref{fig1} and
Fig.\ref{fig2}. One of them is coupled with following the even and odd parts of a photocurrent in the single beam
configuration and the other with two-beam one. In order to characterize our approach it is important in advance to
stress that in distinguish from the other approaches we are going to use the so-called coherent sources, which could
be formally defined by the inequality
\BY
&&\langle\hat E_N(\vec r)\rangle\gg\delta\hat E_N(\vec r,t),\L{1.}
\EY
where on the left  we have an averaged Heisenberg amplitude  and
on the right its fluctuation.   The Heisenberg amplitude is
written in the form
\BY
&&\hat E_N(\vec r,t)=\langle\hat E_N(\vec r)\rangle+\delta\hat E_N(\vec r,t).\L{1.}
\EY
We have a right to say it for both the near and far-fields.
Hereinafter the lower index  $N$ ($F$) means the near- (far-)
field.

 In
the limit of a quasi-monochromatic and quasi-plane travelling wave it is convenient to present it in the form
\cite{Kolobov}
\BY
&&\hat E_N(z,\vec\rho,t)=
i\sqrt{\frac{\hbar\omega_0}{2\varepsilon_0c }}\;e^{\ds
ik_0z-i\omega_0t}\;\hat S_N(\vec\rho,t).\L{1}
\EY
The operator $\hat S_N(\vec\rho,t)$ is normalized amplitude such that the value $\langle\hat S^\dag\hat S\rangle$
takes sense of the mean photon number per $sec$ and per $cm^2$. Under the propagation in free space the normalized
amplitude   obeys the canonical commutation relations
\BY
&&\[\hat S_N(\vec\rho,t),\hat
S_{N}^\dag(\vec\rho^\prime,t^\prime)\]=\delta^2(\vec\rho-\vec\rho^\prime)\;\delta(t-t^\prime),\qquad\[\hat
S_N(\vec\rho,t),\hat S_N(\vec\rho^\prime,t^\prime)\]=0.\L{2}
\EY
Furthermore, we shall assume that the emission from our sources
possess a cylindrical symmetry in the near- and far-field:
\BY
&&\langle\hat S(\vec\rho)\rangle=\langle\hat S(-\vec\rho)\rangle.
\EY
By choosing such experimental situation, where a lens is placed at the focal distances $f$ between the source and the
detection plane the field in front of the detection plane may be considered as a far-field \cite{1}. We shall denote
the corresponding normalized amplitude $\hat S_F(\vec\rho,t)$, hence the far- and near-fields turn out to be coupled
by the integral
\BY
&&\hat S_F(\vec\rho,t)=-\frac{i}{\lambda f}\int d^2\rho^\prime\hat S_N (\vec\rho^\prime,t)\;e^{\ds -i\vec
Q\vec\rho^\prime}=-\frac{2\pi i}{\lambda f}\;\hat S_{N }(\vec Q,t),\qquad \vec Q=\frac{2\pi}{\lambda
f}\;\vec\rho.\L{5.}
\EY
Hereinafter we shall label $G(\vec q)$ as the Fourier image  of a function $G(\vec\rho)$, i.e.,
\BY
&&G(\vec q)=\frac{1}{2\pi}\int d^2\rho\; G (\vec\rho)\;e^{\ds -i\vec q\vec\rho} \qquad\mbox{and}\qquad
G(\vec\rho)=\frac{1}{2\pi}\int d^2q\; G(\vec q) \;e^{\ds i\vec q\vec\rho}.
\EY
In particular in Eq.~(\ref{5.}) we have expressed the Heisenberg far-field amplitude in physical space via the
Heisenberg near-field amplitude in the Fourier domain with $\vec q\to\vec Q$.

Let us remind how the wanted quadrature component is selected in
the HD technique in the near-field. The photocurrent fluctuation
operator $ \delta \hat i =\hat i -\langle\hat i_N \rangle$ reads
 \BY
&&\delta \hat i_N(\vec \rho,t)=\beta(\vec \rho )\; \delta\hat
S_N(\vec \rho,t) +\beta^\ast\delta \hat S_N^\dag(\vec \rho,t),\L{14}
\EY
where $\beta$ is the local oscillator complex amplitude. By choosing the acceptable $\beta$   we can select any
quadrature component at the detector. For the coherent source it is possible partially to avoid the use of the HD
technique because, in this case, the mean amplitude $\langle\hat S_N(\vec \rho,t)\rangle$ takes a role of the local
oscillator amplitude. The amplitude quadrature turns out to be automatically selected but the phase one remains
inaccessible.

If our measurement procedure is replaced to the far-field, then we shall illustrate that there it is possible to find
an acceptable approach when the phase quadrature is selected automatically as well as the amplitude one.

 In the far-field the photocurrent fluctuations  for the coherent  sources are given by
\BY
&&\delta \hat i_F(\vec\rho,t)=\langle\hat
S_F(\vec\rho)\rangle^\ast\;\delta\hat S_F(\vec\rho,t)+\langle\hat
S_F(\vec\rho)\rangle\;\delta\hat S_F^\dag(\vec\rho,t).\L{9}
\EY
Taking into account (\ref{5.}) one can obtain
 \BY
&&\delta \hat i_F(\vec \rho,t)=\(\frac{2\pi}{\lambda f}\)^2\[\langle \hat S_N(\vec Q )\rangle^\ast \;\delta \hat
S_N(\vec Q,t) +\langle \hat S_N(\vec Q )\rangle\delta \hat S_N^\dag(\vec Q,t)\].\L{14}
\EY
Furthermore we shall use the simpler physical situation, where the
specific phases are chosen for our sources. In this section we
require the mean field amplitude in the near field is real,
\BY
&&\langle\hat S_N(\vec\rho,t)\rangle=\langle\hat
S_N(\vec\rho,t)\rangle^\ast.
\EY
In the next section we shall consider the scheme with two sources and for one source we survive the requirement of
the reality but for the second one we shall foresee the phase shift equal to $\pi/2$, i.e., our requirement for the
second source will be $\langle\hat S_N(\vec\rho,t)\rangle=-\langle\hat S_N(\vec\rho,t)\rangle^\ast$. Then
Eq.~(\ref{9}) can be rewritten in the form
 \BY
&&\delta \hat i_F(\vec \rho,t)=\(\frac{2\pi}{\lambda f}\)^2\langle \hat S_N(\vec Q )\rangle\[ \;\delta \hat S_N(\vec
Q,t) +h.c.\].\L{14.}
\EY
Let us introduce the quadrature components $\hat X_N(\vec\rho,t)$ and $\hat Y_N(\vec\rho,t)$ for the near-field
\BY
&&\hat S_N(\vec\rho,t)=\hat X_N(\vec\rho,t)+i\;\hat Y_N(\vec\rho,t),\L{8}
\EY
and correspondingly in the Fourier domain
\BY
&&\hat S_N(\vec Q,t)=\hat X_N(\vec  Q,t)+i\;\hat Y_N(\vec Q,t).\L{8}
\EY
Under the assumption
\BY
&&\langle \hat S_N(\vec Q )\rangle=\langle \hat S_N(\vec Q )\rangle^\ast=\langle\hat X_N(\vec  Q,t)\rangle,
\EY
one can see  both quadratures are presented equally in current
(\ref{14.}).

Let us vary the direct measurement procedure by investigating not the photocurrent itself but its even and odd parts
independently. The even an odd parts   are proportional to summarized and differential photocurrent respectively:
 \BY
&&\delta \hat i_\pm(\vec \rho,t)=\delta \hat i_F(\vec
\rho,t)\pm\delta \hat i_F(-\vec \rho,t).\L{15}
\EY
Now one can get
 \BY
&&\delta \hat i_+(\vec \rho,t)=2\(\frac{2\pi}{\lambda f}\)^2\langle
\hat X_N(\vec Q )\rangle\[\delta \hat X_N(\vec Q,t) +h.c.\],\L{16}\\
&&\delta \hat i_-(\vec \rho,t)=2\(\frac{2\pi}{\lambda f}\)^2\langle \hat X_N(\vec Q )\rangle\;i\!\[\delta \hat
Y_N(\vec Q,t)- h.c.\].\L{17}
\EY
One can conclude that the summarized  photocurrent is determined
by real part of the X-quadrature and the differential photocurrent
- by imaginary part of the Y-quadrature.  On the one hand, it is
not the same result that could be obtained in the HD technique,
where it is possible to select the quadratures  as whole. On the
other hand, in principle, the mathematics provides us with a
possibility to renew the full analytical function by knowing only
its real or imaginary part (the Cauchy-Riemann theorem). Although
this development seems as quite encouraging, nevertheless in the
next section we want to discuss another measurement procedure,
where the quadratures are selected as whole. It is more convenient
because does not require any additional mathematical treatment.

\section{Direct detection of the quadrature components in the far-field (scheme with two coherent sources)\L{III}}
\begin{figure}
% \centering
\includegraphics[height=4.5cm]{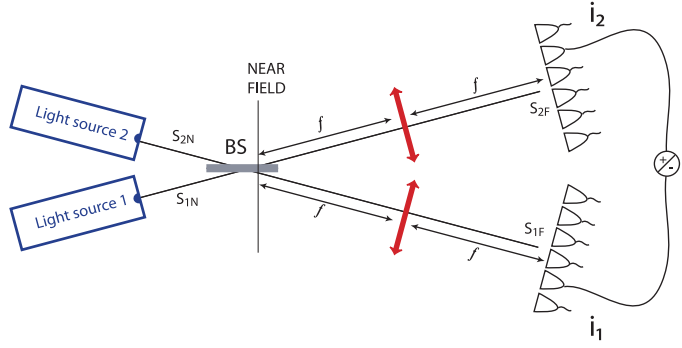}
 \caption{Scheme of a direct detection of two light entangled beams, BS is a beamsplitter, $f$ is a focal distance.}
 \label{fig2}        % Give a unique label
\end{figure}
Let two independent coherent sources emit two beams (see fig.\ref{fig2}) presented correspondingly by the slow
Heisenberg amplitudes $\hat S_{1N}(\vec\rho,t)$ and $\hat S_{2N}(\vec\rho,t)$ in the near-field. As usual the
corresponding quadratures are introduced by relations
\BY
&&\delta\hat S_{mN}(\vec Q,t)=\delta\hat X_{mN}(\vec \rho,t)+i\delta\hat Y_{mN}(\vec \rho,t),\qquad
m=1,2.\L{20}
\EY
As for our coherent sources we assume that they are perfectly identical but there is a phase shift between them equal
to $\pi/2$. In particular this means that the following equalities can take place
\BY
&&\langle\hat S_{1N}(\vec \rho)\rangle=\langle\hat X_{1N}(\vec
\rho)\rangle,\qquad\langle\hat S_{2N}(\vec
\rho)\rangle=i\langle\hat Y_{2N}(\vec \rho)\rangle,\qquad \langle
Y_{2N}(\vec \rho)\rangle=\langle\hat X_{1N}(\vec \rho)\rangle,
\EY
and
\BY
&&\langle\delta\hat X_{1N}(\vec \rho,t)\;\delta\hat X_{1N}(\vec \rho^\prime,t^\prime)\rangle=\langle\delta\hat
Y_{2N}(\vec \rho,t)\;\delta\hat Y_{2N}(\vec \rho^\prime,t^\prime)\rangle,\nn\\
&&\langle\delta\hat Y_{1N}(\vec \rho,t)\;\delta\hat Y_{1N}(\vec \rho^\prime,t^\prime)\rangle=\langle\delta\hat
X_{2N}(\vec \rho,t)\;\delta\hat X_{2N}(\vec \rho^\prime,t^\prime)\rangle.\L{22}
\EY
After mixing in the near-field on the symmetrical beamsplitter we get two other beams with amplitudes
\BY
&&\hat E_{1N}(\vec\rho,t)=\frac{1}{\sqrt2}(\hat S_{1N}(\vec\rho,t)+\hat
S_{2N}(\vec\rho,t)),\qquad\hat E_{2N}(\vec\rho,t)=\frac{1}{\sqrt2}(\hat S_{1N}(\vec\rho,t)-\hat
S_{2N}(\vec\rho,t)).\L{23}
\EY
These beams  can turn out to be entangled if the initial beams were squeezed in the mutually orthogonal quadratures.

Now we again  follow the summarized and differential currents in
the far-field that are defined as
\BY
&&\delta \hat i_\pm(\vec\rho,t)=\delta \hat i_{1F}(\vec\rho,t)\pm\delta \hat i_{2F}(-\vec\rho,t),\L{24}
\EY
where $i_{1F}$ and $i_{2F}$ are correspondingly the currents on
the first and the second pointlike detectors. After non-difficult
algebraic operations one can obtain
\BY
&&\delta \hat i_\pm(\vec\rho,t)=\(\frac{2\pi}{\lambda f}\)^2 \langle X_{1N}(\vec
Q)\rangle\left(\begin{array}{c}(1-i)\[\delta \hat X_{1N}(\vec Q,t)+i\;\delta \hat Y_{2N}(\vec
Q,t)\]+h.c\\\\(1-i)\[\delta \hat X_{2N}(\vec Q,t)+i\;\delta \hat Y_{1N}(\vec Q,t)\]+h.c\end{array}\right).\L{25}
\EY
Let us calculate the correlation functions
\BY
&&\langle\delta \hat i_{\pm}(\vec\rho,t)\;\delta \hat
i_{\pm}(\vec\rho^\prime,t^\prime)\rangle=\nn\\
&&= 4\(\frac{2\pi}{\lambda f}\)^4\langle X_{1N}(\vec Q)\rangle\langle X_{1N}(\vec
Q^\prime)\rangle\[\delta(t-t^\prime)\;\delta^2(\vec Q-\vec Q^\prime)+ 4\left(\begin{array}{c}\langle:\delta \hat
X_{1N}(\vec Q,t)\;\delta \hat X_{1N}(\vec Q^\prime,t^\prime):\rangle\\\\\langle:\delta \hat Y_{1N}(\vec Q,t)\;\delta
\hat Y_{1N}(\vec Q^\prime,t^\prime):\rangle\end{array}\right)\].\L{27.}
\EY
The notation $\langle:\cdots:\rangle $ means the normally ordered
averaging. Under deriving it we have used Eqs.~(\ref{22}) staying
identification of the CS.

From  Eqs.~(\ref{27.}) one can conclude that in the measuring scheme presented in fig.2 it is possible to select any
quadrature component by only choosing the summarized or differential current. We shall call this direct measuring
procedure the {\it plus-minus detection}.

\section{The {\it plus-minus detection} scheme on the basis of two
DOPO or two SPL \L{IV}}
In the previous sections  our requirements relative to sources
were as follows: they are coherent and cylindrically symmetrical.
Bellow we consider the specific situations, namely in this section
we discuss the {\it plus-minus detection} on the basis of two
phase-locked SPL or two DOPO.

The quasi-monochromatic radiation with bounded aperture inside the
cavity at the output mirror  can be formally described by the
Heisenberg amplitude
\BY
&&\hat E_s(z=L, \vec\rho,t)=
i\sqrt{\frac{\hbar\omega_0}{2\varepsilon_0L }}\;e^{\ds
-i\omega_0t}f(\vec \rho)\;\hat a(t).\L{29.}
\EY
Here $\vec\rho$ is the transverse spatial vector, $z$ is a
longitudinal co-ordinate, $L$ is the cavity perimeter, $\omega_0$
is the mode frequency. The normalized amplitudes $\hat a(t)$ and
$\hat a^\dag(t)$ obey the canonical commutation relation
\BY
&&\[\hat a(t),\hat a^\dag(t)\]=1\L{23}
\EY
and $\langle\hat a^\dag(t)\hat a(t)\rangle$ gives the mean photon
number inside the cavity.

The transverse size of the mode is formally bounded by the function
$f(\vec \rho)$ that is normalized such that
\BY
&&\int\;d^2\rho\;|f(\vec \rho)|^2=1,\qquad d^2\rho=dx\;dy.\L{20}
\EY
For the sake of simplicity we shall discuss here the Gaussian
mode, then
\BY
&&f(\vec\rho)=\frac{1}{\sqrt{\pi/2 \;w_0^2}}\;\exp\(
-\frac{\rho^2}{w_0^2}\),\qquad\rho^2=x^2+y^2.\L{31.}
\EY
The value $w_0$ takes a sense of the light spot size.

The Heisenberg amplitude $\hat a(t)$ describes the intracavity field. Certainly, its time evolution
  depends on the specific physical processes inside the cavity. For the coherent field
  \BY
  && \hat a(t)=\langle \hat a\rangle+ \delta\hat a(t),\qquad
\langle \hat a\rangle\gg \delta\hat a.
\EY
The near-field at the output of the cavity with amplitude $\hat
E_{N}(\vec\rho,t)$ is formed as a linear combination of the field
leaving the cavity for a propagation in free space and the
multi-mode vacuum field reflected from the output mirror outside the
cavity
\BY
&&\hat E_{N}(\vec\rho,t)=\sqrt T\;\hat E_s(z=L,\vec\rho,t)-\sqrt R\;\hat
E_{vac}(z=-0,\vec\rho,t).\L{35}
\EY
Here $R$ and $T$ are correspondingly the reflection and transmission
coefficients. We put there are no losses of the field under passing
the mirror, i.e., R+T=1.

Taking into account (\ref{1}) and (\ref{29.}), we can rewrite
(\ref{35}) via the normalized amplitudes
 in the form
\BY
&&\hat S_N(\hat\rho,t)=\sqrt{\kappa }f(\vec\rho)\;\hat a(t)-\hat S_{vac}(\vec\rho,t),\L{36}
\EY
where
\BY
&&\kappa= \frac{cT}{L}\quad\mbox{and}\quad R\approx1,\L{37}
\EY
Here the normalized amplitude $\hat S_{vac}(\vec\rho,t)$ as well as $\hat S_N(\vec\rho,t)$ obeys free-space
commutation relations (\ref{2}). In order to ensure this we should put
\BY
&&\sqrt\kappa \[\hat a(t),\hat a^\dag(t^\prime)\]=\[\hat a(t),\hat
S_{vac}^\dag(t^\prime)\]+\[\hat S_{vac}(t),\hat
a^\dag(t^\prime)\]. \L{27}
\EY
Because we want to consider the $\pm$detection we need to foresee
two sources and correspondingly we have to discuss the amplitudes
$\hat S_{r,vac}(\vec\rho,t)$, $\hat S_{r,N}(\vec\rho,t)$, and
$\hat a_{r}( t)$, where index $r=1,2$ indicates the source. We
shall assume that both sources are perfectly identical with each
other, and the parameters $\kappa$ and $f(\vec\rho)$ are the same
for both. However it is suggested that there is the phase shift
between them such that
\BY
&&\langle\hat a_{1} \rangle=\sqrt{ n},\qquad \langle\hat a_{2}
\rangle=i\sqrt{ n}
 ,\L{}
\EY
and  equalities (\ref{22}) take place.

Applying these considerations to our sources, it is not difficult to get the quadrature components
\BY
&&\delta \hat X_{r,N}(\vec\rho,t)=\sqrt{\kappa } \; f(\vec\rho
)\delta\hat x_r (t) -\hat {
X}_{r,vac}(\vec\rho,t),\L{}\\
&& \delta\hat Y_{r,N}(\vec\rho,t)=\sqrt{\kappa  }  \;f(\vec\rho
)\delta\hat y_r (t) -\hat {
Y}_{r,vac}(\vec\rho,t),\L{}\\
&&\langle\hat X_{1,N}(\vec\rho)\rangle=\langle\hat Y_{2,N}(\vec\rho)\rangle=\sqrt{\kappa n} \; f(\vec\rho ),\L{}\\
&&\langle\hat X_{2,N}(\vec\rho)\rangle=\langle\hat
Y_{1,N}(\vec\rho)\rangle  =0,
 \EY
where $\delta\hat x_r(t)$ and $ \delta\hat y_r(t)$ are the quadrature components of the intracavity field for both
sources
\BY
&&\delta \hat a_r(t)=\delta\hat x_r(t)+i \delta\hat y_r(t).\qquad r=1,2.
\EY
Passing on to the Fourier domain, one can obtain
\BY
&&\delta \hat x_{i,N}(\vec Q,t)=\sqrt{\kappa }\; f_{\vec Q}\delta\hat x_i(t) -\hat {
x}_{i,vac}(\vec Q,t),\L{45}\\
&& \delta\hat y_{i,N}(\vec Q,t)=\sqrt{\kappa  }\; f_{\vec Q}\; \delta\hat y_i(t) -\hat {
y}_{i,vac}(\vec Q,t),\L{}\\
&&\langle\hat x_{1,N}(\vec Q)\rangle=\langle\hat y_{2,N}(\vec Q)\rangle=\sqrt{\kappa n}\; f_{\vec
Q}.\L{47}
 \EY
Here
\BY
&&f_{\vec Q}=\frac{1}{2\pi}\int d^2\rho f(\vec\rho)e^{\ds -i\vec Q\vec\rho}
\EY
and for the Gaussian mode (\ref{31.})
\BY
&&f_{\vec Q}=\sqrt{\frac{w_0^2}{2\pi}}e^{\ds - \frac{1}{4}w_0^2Q^2}=\sqrt{\frac{w_0^2}{2\pi}}e^{\ds
-\frac{1}{4}\frac{ \rho^2}{\tilde w_0^2}},\qquad\tilde w_0=\frac{\lambda f}{2\pi}\;\frac{1}{w_0}.
\EY
Now we have a complete information about our sources to calculate
a final correlation function. Substituting (\ref{45})-(\ref{47})
to Eq.~(\ref{27.}), one can obtain
\BY
&&\langle\delta \hat i_{\pm}(\vec\rho,t)\;\delta \hat i_{\pm}(\vec\rho^\prime,t^\prime)\rangle= 4\kappa n
\(\frac{2\pi}{\lambda f}\)^4f_{\vec Q}f_{\vec Q^\prime}\[\delta(t-t^\prime)\;\delta^2(\vec Q-\vec Q^\prime)+
4\kappa\left(\begin{array}{c}\langle:\delta \hat x_{1}(t)\delta \hat x_{1}(t^\prime):\rangle\\\\\langle:\delta \hat
y_{1}(t)\delta \hat y_{1}(t^\prime):\rangle\end{array}\right)f_{\vec Q}f_{\vec Q^\prime}\].\L{46.}
\EY
Let us consider the spectrum of the spatio-temporal correlation
function determined as
\BY
&&(\delta \hat i_\pm^2)_{\vec q,\Omega}= \lim_{T\to\infty}\frac{1}{T}\int^{T/2}_{-T/2}\int^{T/2}_{-T/2}
dt\;dt^\prime\int\int d^2\rho\; d^2\rho^\prime\langle\delta \hat i_\pm(\vec\rho,t)\;\delta \hat i_\pm(\vec
\rho^\prime,t^\prime)\rangle\;e^{\ds i\Omega(t-t^\prime)}e^{\ds i\vec q(\vec\rho-\vec\rho^\prime)}.\L{27..}
\EY
Substituting (\ref{46.})  to (\ref{27..}) after all integrations
one can obtain the spectra in the explicit form. For DOPO it can
be read via the physical parameters:
\BY
&&(\delta \hat i_\pm^2)_{\vec q,\Omega}= 4\kappa n \[1\pm   e^{\ds
-\tilde w_0^2q^2}\left(\begin{array}{c}\ds
\frac{\kappa^2}{\kappa^2(\mu_{th}-1)^2+\Omega^2}\\\\\ds\frac{\kappa^2}{\kappa^2\mu_{th}^2+\Omega^2}\end{array}\right)
\],\qquad
\tilde w_0=\frac{\lambda f}{2\pi}\;\frac{1}{w_0} .\L{52.}
\EY
Under calculations of these formulas we have used the expressions
for the correlation functions $\langle:\delta \hat x_{1}(t)\delta
\hat x_{1}(t^\prime):\rangle$ and $\langle:\delta \hat
y_{1}(t)\delta \hat y_{1}(t^\prime):\rangle$ obtained in
Ref.~\cite{6}. Here $\mu_{th}>1$ is a factor determining a
relationship between the pump power and the threshold one.

By the same way we could find the spectra for phase-locked SPLs in
the scheme of {\it plus-minus detection}. Using the results
obtained in Ref.~\cite{2,2.} one can read
\BY
&&(\delta \hat i_{\pm}^2)_{\vec q,\Omega}= 4\kappa n \[1\mp
e^{\ds -\tilde w_0^2q^2} \left(\begin{array}{c}\ds
\frac{\kappa^2}{\kappa^2+\Omega^2}\\\\\ds\frac{2\kappa^2}{\kappa^2/4+\Omega^2}\end{array}\right)\].\L{55.}
\EY
It is not difficult to check that the measurements in the HD technique in the near-field will give exactly the same
results. The last formulas exhibit the possibility of the {\it plus-minus detection} in the cases of the simplest
single-mode sources. One can see that it is possible to follow both the squeezed and stretched quadratures
simultaneously without the use of the HD technique.

\section{Pixellised source: array of the coherent pointlike sources\L{V}}
Let us consider more complicated sources that we call the PS. Let $N^2$  pointlike identical sources be placed
periodically on the plane such that the linear distance between the adjoining pixels is much more than the beam spot
each of the individual pixels
\BY
&&l\gg w_0.
\EY
This requirement allows us to say that all sources are perfectly independent from each other. Simultaneously we
assume that all the lasers or DOPO used as the pixels  are perfectly synchronized by means of the pump field for the
DOPO or by means of the special synchronizing external field for the lasers \cite{2,2.}.

In order to construct theoretically the full beam from the PS we
can use the linear superposition principle and derive the full
Heisenberg amplitude  as a sum of the individual amplitudes from
each of the pixels,
\BY
&&\hat E(\vec r,t)=\sum_m\hat E_m(\vec r,t).\L{ }
\EY
The positions of the pixels on the surface is determined by the
vectors
\BY
&&\vec\rho_m=l\;\vec m,\qquad m_x,m_y=0,\pm1,\pm2,\ldots,\pm
(N-1)/2.\L{28}
\EY
In Eq.~(\ref{36}) we have written the explicit form for the slow Heisenberg amplitude in the case of the single
pixel. It is not difficult to generalize it on the case of the PS
\BY
&&\hat S(\vec\rho,t)=\sqrt{\kappa }\sum_{\vec\rho_m}
f(\vec\rho-\vec\rho_m)\;\hat a_{m}(t) -\hat {
S}_{vac}(\vec\rho,t).\L{29}
\EY
The function $f(\vec\rho-\vec\rho_m)$ describes the $m$-beam
aperture. Since we choose $l\gg w_0$, then all the pixels are
independent from each other and we have a right to require that
\BY
&&\int  |f(\vec\rho-\vec\rho_m)|^2d^2\rho=1
\EY
and
 \BY
&& \[\hat a_{m}(t),\hat a_{n}^\dag(t)\]=\delta_{mn}.
\EY

The operators $\hat a_m$ are the normalized amplitudes describing a single mode field inside each of the individual
cavities.

As before we consider the coherent fields that is
\BY
&&\hat a_{m}(t)=\langle \hat a_{m}\rangle+\delta\hat
a_{m}(t),\qquad\langle \hat a_{m}\rangle\gg\delta\hat
a_{m}(t).\L{32}
\EY
Because under {\it plus-minus detection} we have to foresee two
independent PSs we are under obligation to specify belonging the
$m$-th pixel to one of these PSs that is to add additional index:
$\hat a_m\to\hat a_{r,m}$, where $r=1,2$.

 Now
it is not difficult to generalize formulas (\ref{45})-(\ref{47})
\BY
&&\delta \hat X_{r,N}(\vec Q,t)=\sqrt{\kappa }\; f_{\vec Q}\sum_{\vec\rho_m}e^{-i\vec Q\vec\rho_m}\delta\hat
x_{r,m}(t) -\hat { X}_{r,vac}(\vec Q,
t),\L{62}\\
&& \delta\hat Y_{r,N}(\vec Q,t)=\sqrt{\kappa  }\; f_{\vec Q}\sum_{\vec\rho_m}e^{-i\vec Q\vec\rho_m}\delta\hat
y_{r,m}(t) -\hat {
Y}_{r,vac}(\vec Q,t),\L{}\\
&&\langle\hat X_{1,N}(\vec Q)\rangle=\langle\hat Y_{2,N}(\vec Q)\rangle=\sqrt{\kappa n}\; f_{\vec Q}\Lambda_{\vec Q}
.\L{64}
 \EY
Here $\delta\hat x_{r,m}(t)$ and $ \delta\hat y_{r,m}(t)$ are the quadrature components for the intracavity
single-mode field
\BY
&&\delta \hat a_{r,m}=\delta\hat x_{r,m}(t)+i \delta\hat y_{r,m}(t)
\EY
and
\BY
&&\Lambda_{\vec Q}=\sum_{\vec\rho_m}e^{-i\vec
Q\vec\rho_m}.\L{Lambda}
\EY
Evaluating the series in Eqs.~(\ref{Lambda}) one can get it in the
explicit form,
\BY
&&\Lambda_{\vec Q}=\frac{\sin Q_xlN/2}{\sin Q_xl/2}\;\frac{\sin
Q_ylN/2}{\sin Q_yl/2},
\EY
where $Q_x$ and $Q_y$ are the components of vector $\vec Q$. On
the basis of Eqs.~(\ref{62})-(\ref{64}) the correlation functions
can be read
\BY
&&\langle\delta \hat i_{\pm}(\vec\rho,t)\;\delta \hat i_{\pm}(\vec\rho^\prime,t^\prime)\rangle=\nn\\&&\nn\\
&&=4\kappa n\(\frac{2\pi}{\lambda f}\)^4f_{\vec Q}\Lambda_{\vec Q}\;f_{\vec Q^\prime}\Lambda_{\vec
Q^\prime}\[\delta(t-t^\prime)\delta^2(\vec Q-\vec Q^\prime)+4\kappa
\left(\begin{array}{c}\langle:\delta\hat x_r(t)\;\delta\hat
x_r(t^\prime):\rangle\\\\\langle:\delta\hat y_r(t)\;\delta\hat
y_r(t^\prime):\rangle\end{array}\right)\;f_{\vec Q}\;f_{\vec Q^\prime}\Lambda_{\vec Q-\vec
Q^\prime}\].
\EY
Passing to the Fourier domain, one can obtain again for the DOPO
and SPL Eqs.~(\ref{52.})-(\ref{55.}), where we should do the
following exchange
\BY
&&e^{\ds -\tilde w_0^2\vec q^2}\quad\to\quad\frac{1}{N^2}\sum_{\vec Q_m,\vec Q_n} e^{\ds -\tilde w_0^2(\vec q+\vec
Q_m-\vec Q_n)^2},\qquad\vec Q_m=\frac{2\pi}{\lambda f}\vec\rho_m.
\EY
\begin{figure}
% \centering
\includegraphics[height=4.5cm]{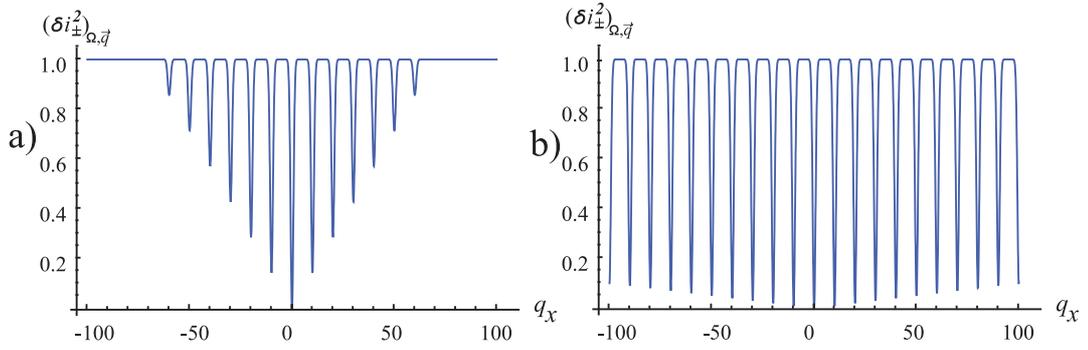}
 \caption{Spatial spectrum of the summarized (for SPL) and  differential (for DOPO) photocurrent
 variances; $q_y=0, \Omega=0, l/w_0=10$ in arbitrary units, $N=7$ (a), $N=99$ (b).}
 \label{fig3ab}        % Give a unique label
\end{figure}

In Fig. \ref{fig3ab} one can see that the photocurrent spectrum
exhibits the periodical structure in dependence on spatial
frequency $q$ that reflects directly the periodical geometry of
the PS. The reduction of the shot noise is more effective for
bigger N, where the bigger number of holes with nearly perfect
noise reduction appears.

\section{Direct detection in the single beam \L{VI}}
Let us return to the measuring procedure presented in Fig.\ref{fig1}, where we follow the even and odd parts of
current under the detection in the beam from the single PS.

Taking into account the formulas from the previous section it is possible to get the explicit expression for the pair
correlation function of the summarized and differential currents in the form
\BY
&&\langle\delta \hat i_{\pm}(\vec\rho,t)\;\delta \hat
i_{\pm}(\vec\rho^\prime,t^\prime)\rangle=2\kappa
n\(\frac{2\pi}{\lambda f}\)^4f_{\vec Q}\Lambda_{\vec Q}\;f_{\vec
Q^\prime}\Lambda_{\vec Q^\prime}\[\delta(t-t^\prime)\(\delta^2(\vec
Q-\vec Q^\prime)+
\delta^2(\vec Q+\vec Q^\prime)\)+ \right.\nn\\
&&\left.+4\kappa \left(\begin{array}{c}\langle:\delta\hat
x_r(t)\;\delta\hat x_r(t^\prime):\rangle\\\\\langle:\delta\hat
y_r(t)\;\delta\hat y_r(t^\prime):\rangle\end{array}\right)\;f_{\vec
Q}\;f_{\vec Q^\prime}\(\Lambda_{\vec Q+\vec
Q^\prime}\pm\Lambda_{\vec Q-\vec Q^\prime}\)\].
\EY

\begin{figure}
 \centering
\includegraphics[height=4.5cm]{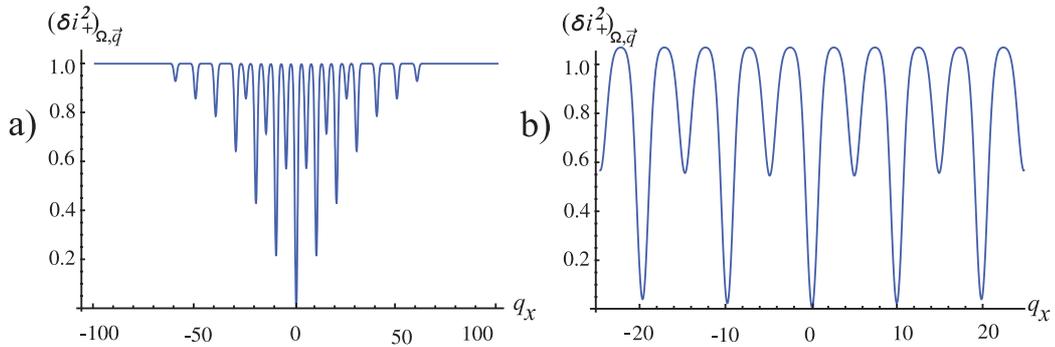}
  \caption{Spatial summarized photocurrent variances spectrum in the far field, in case when
SPL is chosen as the pixel, $q_y=0, \Omega=0, l/w_0=10$ in arbitrary
units, $N=7$ (a), $N=99$ (b).}
 \label{fig2ac}        % Give a unique label
\end{figure}

\begin{figure}
 \centering
\includegraphics[height=4.5cm]{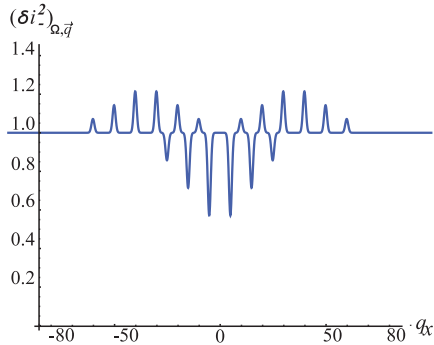}
  \caption{Spatial differential photocurrent variances spectrum in the far field, in case when
DOPO is chosen as the pixel, $q_y=0, \Omega=0, l/w_0=10$ in
arbitrary units, $N=7$.}
 \label{fig2b}        % Give a unique label
\end{figure}

Using again the definition (\ref{27..}) it is possible to get the
spectra for the DOPO and SPL in the form Eqs.~(\ref{52.}) and
(\ref{55.}), where we should do the following exchange
\BY
&&e^{\ds -\tilde w_0^2\vec q^2}\quad\to\quad\frac{1}{2N^2} {\sum_{\vec Q_m\vec Q_n}}
\[e^{\ds -\tilde w_0^2\(\vec q+\frac{1}{2}(\vec Q_m-\vec Q_n)\)^2}\pm
e^{\ds -\tilde w_0^2\(\vec q+\vec Q_m-\vec Q_n\)^2}\] .
\EY
The formula   exhibits a set of resonant Gaussian picks or holes on
the level of the shot noise (see figs.\ref{fig2ac} and \ref{fig2b}).
Each of the picks has a width
\BY
&& \Delta q=\frac{1}{\tilde w_0}=\frac{2\pi}{\lambda f}\;w_0,
\EY
and is centered on the frequency $\vec Q_m$ or $\vec Q_m/2$ that
matches the pixel positions:
\BY
&& \vec Q_m=\frac{2\pi}{\lambda f}\;\vec \rho_m.
\EY
The distance between the adjoining picks along the x- or y-axes $d$ and the full actual frequency range $D$ are given
by
\BY
&&d=\frac{2\pi}{\lambda f}\;l\quad\mbox{and}\quad
D=\frac{2\pi}{\lambda f}\;lN. \L{52}
\EY
Let us remind that values $w_0$, $l$ $(w_0\ll l)$, and $N$,
respectively, corresponds to the linear size of the pixel, a
distance between the adjoining pixels, and the pixel number along
one of the transverse directions.

%pictures
%%

We want to follow the reduction of the shot noise, i.e., the
summarized current in the case of the SPL pixels and the
differential one in the case of the DOPO pixels. In
(fig.\ref{fig2ac}), an interchange for the holes on the main
frequencies (in case $l/w_0=10$ they are frequencies with numbers
divisible by $10$ and zero frequency) and half less deep holes on
the additional frequencies (here with the numbers divisible by $5$)
occurs.

Both sequences of the holes decline to the shot noise level on the
frequencies far from zero. The effective hole number is determined
by the parameter D (Eq. (\ref{52})). One can see that for small
$N$ (see fig.\ref{fig2ac}$a$), complete reduction of the shot
noise takes place only around zero frequency $q_x$. For the
greater number of $N$ (see fig.\ref{fig2ac}$b$), the interval
which damping of the holes in the shot noise level occurs on gets
larger. Therefore  the well reduced fluctuations can be observed
at the wider range of $q_x$.

The situation when  the DOPO is chosen as the pixels
(fig.\ref{fig2b}) is unfavorable since deeps at the shot noise level
dependency are alternating with picks. Moreover, the depth of the
holes is limited by half of shot noise level. Therefore the perfect
suppression of the shot noise at this case is simply impossible.

\section{conclusion\L{VII}}
In this article we purpose two main aims. First of all we wanted
to find the measuring procedure that could be in some cases an
alternative to the balanced homodyne detection. The latter is an
universal approach, which allows to select any quadrature
component of the field by mixing on the detector with the field
from the local oscillator. For that we have to choose the
acceptable phase and the acceptable spatio-temporal configuration
of the local oscillator amplitude. Often especially in quantum
imaging, where the spatial configuration can be extremely
complicated, this task turns out to be not very simple. Then it
would be nice to find an approach, where these obstacles could be
graded to some way.

To our mind   we have discussed the quite adequate scheme that
allows to follow both squeezed and stretched quadrature components
simultaneously without the use of the HD technique. For that we need
first to use the so-called coherent sources and second to follow not
current itself but the summarized and differential combinations
currents under the detection of two entangled beams in the
far-field.

The second goal was as follows. We constructed the theory of
so-called pixellised source of the spatio-temporal squeezed light.
We assumed to use the phase-locked sub-Poissonian lasers and the
optical parametric oscillators generating above threshold as
pixels. On this basis we have examined  our measuring scheme.

\section{ACKNOWLEDGMENT}

This study was performed within the framework of the Russian–French Cooperation Program “Lasers and Advanced Optical
Information Technologies” and the European Project HIDEAS (grant No. 221906), and supported by  RFBR (No.
08-02-92504, and No. 08-02-00771).

\end{document}